\documentclass[twocolumn,showpacs,preprintnumbers,amsmath,amssymb]{revtex4}
\usepackage{graphicx}
\usepackage{dcolumn}
\usepackage{bm}

\newcommand{\bef}{\begin{figure}}
\newcommand{\eef}{\end{figure}}
\newcommand{\nn}{\nonumber}
\newcommand{\be}{\begin{equation}}
\newcommand{\ee}{\end{equation}}
\newcommand{\bea}{\begin{eqnarray}}
\newcommand{\eea}{\end{eqnarray}}
\begin{document}

\title{Evolution of collectivity as a signal of quark gluon plasma formation in heavy ion collisions}

\author{Payal Mohanty, Jan-e Alam and  Bedangadas Mohanty}

\medskip

\affiliation{Variable Energy Cyclotron Centre, Kolkata 700064, India}

\date{\today}

\begin{abstract}
A  measurement for studying the mass dependence of the dilepton 
interferometry in relativistic heavy-ion collision experiments as a 
tool to characterize the quark-gluon phase is proposed.  
In calculations involving dileptons, we show that the 
mass dependence of radii extracted from the virtual photon (dilepton) 
interferometry provide access to the development of collective flow with 
time. It is argued that the non-monotonic variation of HBT radii with
invariant mass of the lepton pairs signals the formation of quark gluon plasma
in heavy ion collisions. Our proposal of experimentally measuring the 
ratio, $R_{\mathrm out}/R_{\mathrm side}$ for dileptons can be used 
to estimate the average life times of the partonic as well as the hadronic
phases.
\end{abstract}

\maketitle

\section{INTRODUCTION}

The primary goal of the nuclear collision experiments at ultra-relativistic
energies is to create and study a new  state of matter called the
Quark Gluon Plasma (QGP)~\cite{whitepapers}. 
Most of the experimental observables for QGP, however, contain 
contributions from both the partonic as well as hadronic phases. 
As a consequence the disentanglement of the signals 
for the partonic phase remains a challenge and has been 
successful to a limited extend~\cite{whitepapers}. 
Some of the probes, which are produced early in the interactions 
like jets~\cite{jetquenching} and heavy quarks~\cite{phenixnpe} or 
real  photons~\cite{larry,ja1,ja2}  and dileptons~\cite{gale,rapp,NA60,phenixdil} 
are considered to be particularly useful. 

The transverse  momentum ($p_T$) distribution of
photons reflect the temperature of the source as their productions
from  a thermal system 
depend on the temperature ($T$)
of the bath through the thermal phase space 
factors of the participants of the
reaction that produces the photon.
However, the thermal phase space factor may be changed
by  several factors -  {\it e.g.} the transverse
kick due to flow received by  low $p_T$ photons from the low 
temperature hadronic phase will mingle with 
the high $p_T$ photons from the partonic phase, making
the task of detecting and characterizing QGP more difficult.
For dilepton the situation is, however, different because in this
case we have two kinematic variables - out of these two, the
$p_T$ spectra of lepton pairs is affected 
but the $p_T$ integrated invariant mass ($M$) spectra
is unaltered by the  flow.  
Moreover, it is expected that the large $M$ 
pairs originate from the early time and the low $M$ pairs
predominantly produced in the late time. Therefore,
the $M$ distribution can act as a chronometer of
the heavy ion collisions.
This suggests that a
simultaneous measurement of $p_T$ and $M$ spectra
along with a judicious choice of $p_T$ and 
$M$ windows will be very useful to characterize the
QGP and the hadronic phases. 
Precise measurements of lepton pairs in $pp$ collisions at a given
collision energy is of paramount importance for detecting the thermal
spectra in heavy ion collisions at the same energy. 

Experimental measurements of two-particle intensity interferometry
has been established as a useful tool 
to characterize the space-time evolution of 
the heavy-ion collision~\cite{hb3}. For the case of dileptons,
such an interferometry needs to be carried out over dilepton pairs,
theoretically representing a study of the correlations between two 
virtual photons.  Although, the dilepton production rate is down
by a factor of $\alpha$ compared to real photon, the analysis involving
lepton pairs has been successfully used to get direct photon 
yields at  RHIC~\cite{phenixphoton}. 
In contrast to hadrons, two-particle intensity interferometry by using 
lepton pairs, like photons, have almost no interaction with 
the surrounding hadronic medium hence can provide information on the 
history of the evolution of the hot matter  
very efficiently. From the experimental point of view, 
dilepton interferometry encounters
considerable difficulties compared to hadron interferometry due 
to small yield of the dileptons from the early hot and 
dense region of the matter and the associated large background primarily 
from the electromagnetic decay processes of hadrons at freeze-out. 
However, recent work demonstrate that it is still 
possible to carry out experimentally such interferometry 
studies~\cite{peressounko}. 
With a high statistics data already collected at RHIC in the year 2010 
by both STAR and PHENIX collaborations 
having dedicated detectors (Time-Of-Flight~\cite{tof} 
and Hadron Blind Detector~\cite{hbd}) with good acceptance
for dilepton measurements, also augurs well for the dilepton 
interferometry  analysis.   

In this work we present this new proposal for 
carrying out an experimental measurement of dilepton interferometry 
at RHIC. We establish through a hydrodynamical model based 
space-time evolution 
for central 0-5\% Au+Au collisions at $\sqrt{s_{NN}}$ = 200 GeV 
the promise of such a dilepton interferometry analysis will hold out to
understand the properties of the partonic
phase.

\section{Dilepton interferometry in heavy ion collision}
As interferometry of the dilepton pairs actually reflect correlations
between two virtual photons, the analysis then concentrates on
computing the Bose-Einstein correlation (BEC) function for two 
identical particles defined as,
\begin{equation}
C_{2}(\vec{k_{1}}, \vec{k_{2}}) = \frac{P_{2}(\vec{k_{1}}, \vec{k_{2}})}
{P_{1}(\vec{k_{1}}) P_{1}(\vec{k_{2}})} 
\label{eq1}
\end{equation}
where $\vec{k_i}$ is the three momentum of the particle $i$ and 
$P_{1}(\vec{k_{i}})$
and $P_{2}(\vec{k_{1}}, \vec{k_{2}})$ represent 
the one- and two- particle inclusive
lepton pair spectra respectively. The correlation function, $C_2$
has been evaluated with the following source function:
\begin{equation}
P_{1}(\vec{k}) = \int d^{4}x~\omega (x,k)
\label{eq2}
\end{equation}
and
\begin{widetext}
\begin{equation}
P_{2}(\vec{k_{1}},\vec{k_{2}})= 
P_{1}(\vec{k_{1}})P_{1}(\vec{k_{2}})
+\int d^{4}x_{1} d^{4}x_{2} ~\omega (x_{1},K) 
\omega (x_{2},K)~\cos(\Delta x^{\mu} \Delta k_{\mu}) 
\label{eq3}
\end{equation}
\end{widetext}
where $K=(k_1+k_2)/2$, $\Delta k_\mu=k_{1\mu}-k_{2\mu}=q_\mu$,
$x$ and $k$ are the four co-ordinates for position and momentum 
variables respectively and $\omega(x,k)$ is the source function 
related to the thermal emission rate 
of virtual photons  
per unit four volume, $R$ as follows:
\begin{widetext}
\begin{equation}
\omega(x,k)=\int_{M_1^2}^{M_2^2}\,dM^2\,\left[\frac{dR}{dM^2d^2k_Tdy}\right]
\label{eq2}
\end{equation}
\end{widetext}
The inclusion of the spin of the  virtual photon  
will reduce the value of $C_2-1$ by 1/3. 
The correlation functions can be evaluated for different average 
mass windows,
$\langle M\rangle$ ( $ \equiv M_{e^{+}e^{-}}$)= $(M_1+M_2)/2$.
The leading order process through which lepton pairs are produced in QGP 
is $q\bar{q}\rightarrow l^+l^-$~\cite{qqpair}. For the low $M$ dilepton 
production from the hadronic phase the decays of the light vector mesons 
$\rho, \omega$ and $\phi$ have been considered
including the continuum~\cite{larry,ja1,ja2,gale,rapp,shu}.  
Since the continuum 
part of the vector meson spectral functions are included in the current work 
the processes like four pions annihilation~\cite{4pi} are excluded to avoid 
double counting. 
\begin{table}
\caption{Values of the various parameters used in the relativistic hydrodynamical calculations.}
\label{initialconditions}
\begin{tabular}{|c|c|}
\hline
$T_{i}$ & 290 MeV\\
\hline
$\tau_{i}$& 0.6 fm\\
\hline
$T_{c}$ & 175 MeV\\
\hline
$T_{ch}$ & 170 MeV\\
\hline
$T_{fo}$ & 120 MeV\\
\hline
EoS & 2+1 Lattice QCD~\cite{MILC}\\
\hline
\end{tabular}
\end{table}
For the space time evolution of the system relativistic hydrodynamical
model with cylindrical symmetry~\cite{hvg} and  boost invariance along
the longitudinal direction~\cite{jdb} has been used. The initial 
temperature ($T_{i}$) and proper thermalization time 
($\tau_{i}$) of the system 
is constrained by hadronic multiplicity ($dN/dy$) as $dN/dy\sim T_i^3\tau_i$.
The equation of state (EoS) which controls the rate of expansion/cooling
has been taken from the lattice QCD calculations~\cite{MILC}.
The chemical ($T_{ch}$) and kinetic ($T_{fo}$) freeze-out temperatures 
are fixed by the particle ratios and the slope of the $p_T$ 
spectra of hadron~\cite{hirano}. The values of these parameters are
displayed in Table~\ref{initialconditions}.

With all these ingredients we evaluate the correlation function $C_2$
for different invariant mass 
windows as a function of $q_{side}$ and $q_{out}$ which are related 
to transverse momentum of individual pair as follows:
$q_{out}=(k_{1T}^2-k_{2T}^2)/f(k_{1T},k_{2T})$ 
and $q_{side}=(2 k_{1T}k_{2T}\sqrt{1-\cos^{2}(\psi_{1}-\psi_{2})})/
{f(k_{1T},k_{2T})}$
where $f(k_{1T},k_{2T})=
{\sqrt{k_{1T}^2+k_{2T}^2+2 k_{1T}k_{2T}\cos(\psi_1-\psi_2)}}$,
$k_i=(M_{iT}coshy_i,k_{iT}cos\psi_i,k_{iT}sin\psi_i, M_{iT}sinhy_i$)
and $M_{iT}=\sqrt{k_{iT}^2+<M>^2}$.

\bef
\begin{center}
\includegraphics[scale=0.4]{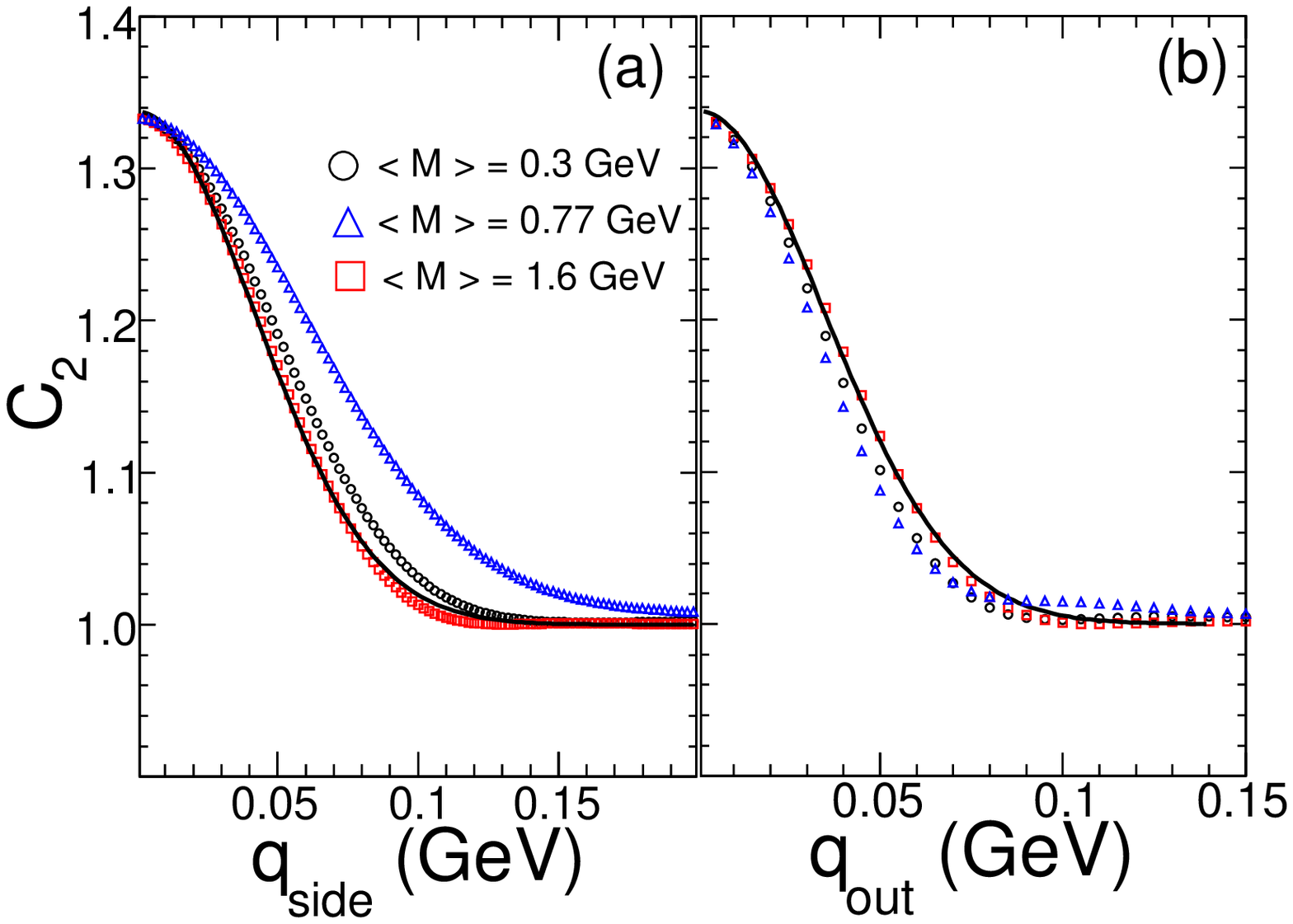}
\caption{(color online) 
Correlation function for dilepton pairs as a function of 
$q_{side}$ [(a), for $k_{1T}=k_{2T}=2$ GeV and $\psi_2=0$] 
and $q_{out}$ [(b), for $\psi_1=\psi_2=0$ and $k_{1T}=2$ GeV] 
for three values of $\langle M\rangle$. The solid lines shows the
parameterization of $C_2$ using Eq.~\ref{ceqn}.}
\label{fig1}
\end{center}
\eef
\section{Results and discussions}
Fig.~\ref{fig1} shows the BEC function of dilepton pairs for different
values of $\langle M\rangle$ as a function of   $q_{\mathrm side}$ and 
$q_{\mathrm out}$.
We have evaluated the $C_2$  for $\langle M \rangle=$0.3, 0.5, 0.77, 1.2, 1.6 
and 2.5 GeV, 
however, in Fig.~\ref{fig1} we display  results for 
only three values of $\langle M\rangle$ corresponding to low and high mass 
which are expected to be dominated by radiations 
(see  Fig.~\ref{fig2}) from QGP ($\langle M\rangle \sim 0.3, 1.6$ 
GeV) and  hadronic phase ($\langle M\rangle\sim 0.77$ GeV)
respectively. 
A clear difference is seen in $C_2$  for different $\langle M\rangle$
windows when plotted as a function of $q_{\mathrm side}$.
The differences are, however, small when BEC is studied as a function of 
$q_{out}$.

The source dimensions can be 
obtained by parameterizing the calculated correlation function 
of the (time like) virtual photon with the empirical (Gaussian) form:
\be
C_2=1+\lambda\exp(-R^2_{i}q^2_{i}).
\label{ceqn}
\ee
where the subscript $i$ stand for $out$ and $side$ and $\lambda$ 
($=1/3$ here) represents
the degree of chaoticity of the source. 
The deviation of $\lambda$ from
1/3 will indicate the presence of non-thermal sources. 
A representative fit to the correlation functions are shown in 
Fig.~\ref{fig1} (solid lines).

\bef
\begin{center}
\includegraphics[scale=0.3]{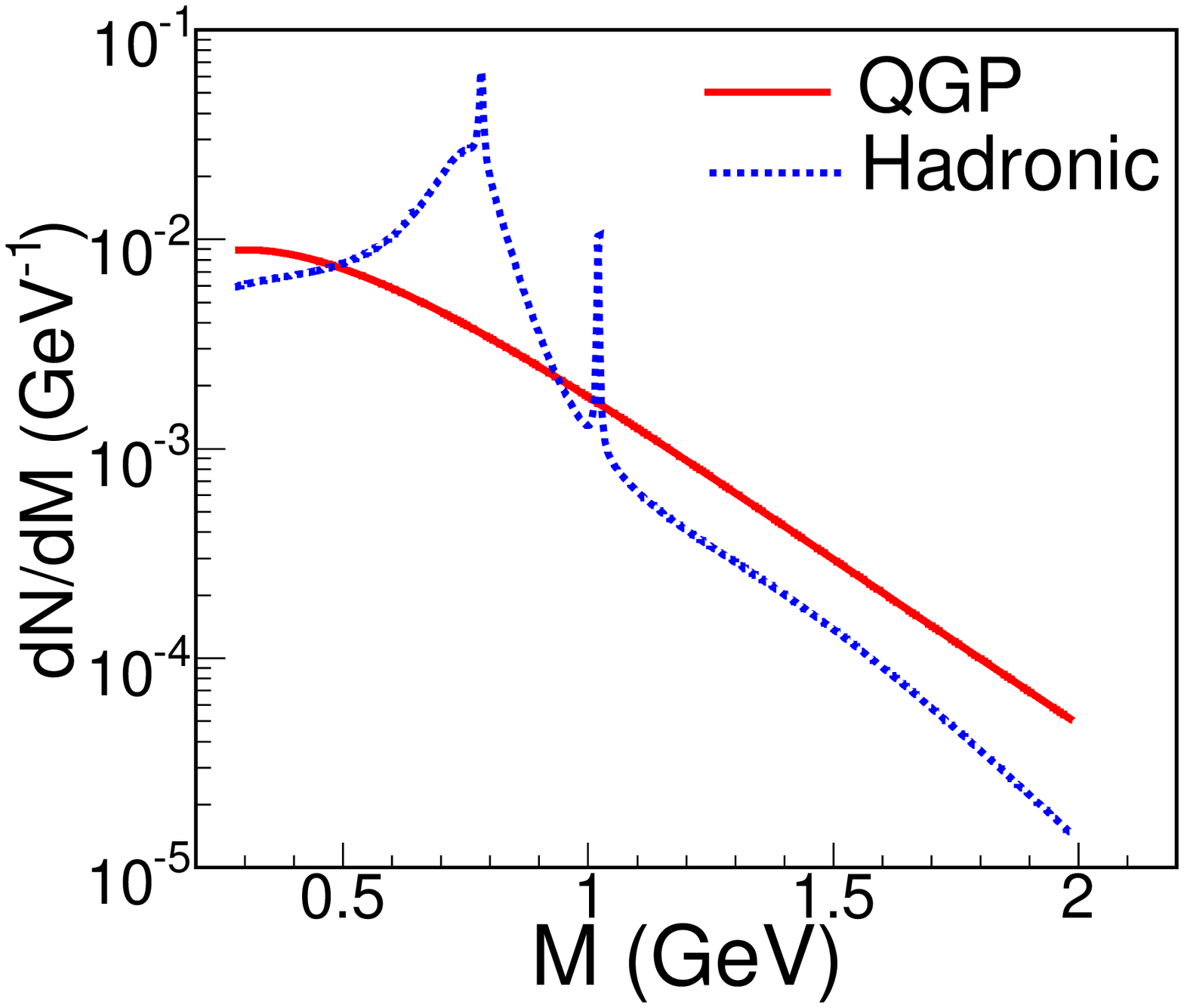}
\caption{(color online) Invariant mass distribution of lepton pairs from quark matter
and hadronic matter.
}
\label{fig2}
\end{center}
\eef

\bef
\begin{center}
\includegraphics[scale=0.3]{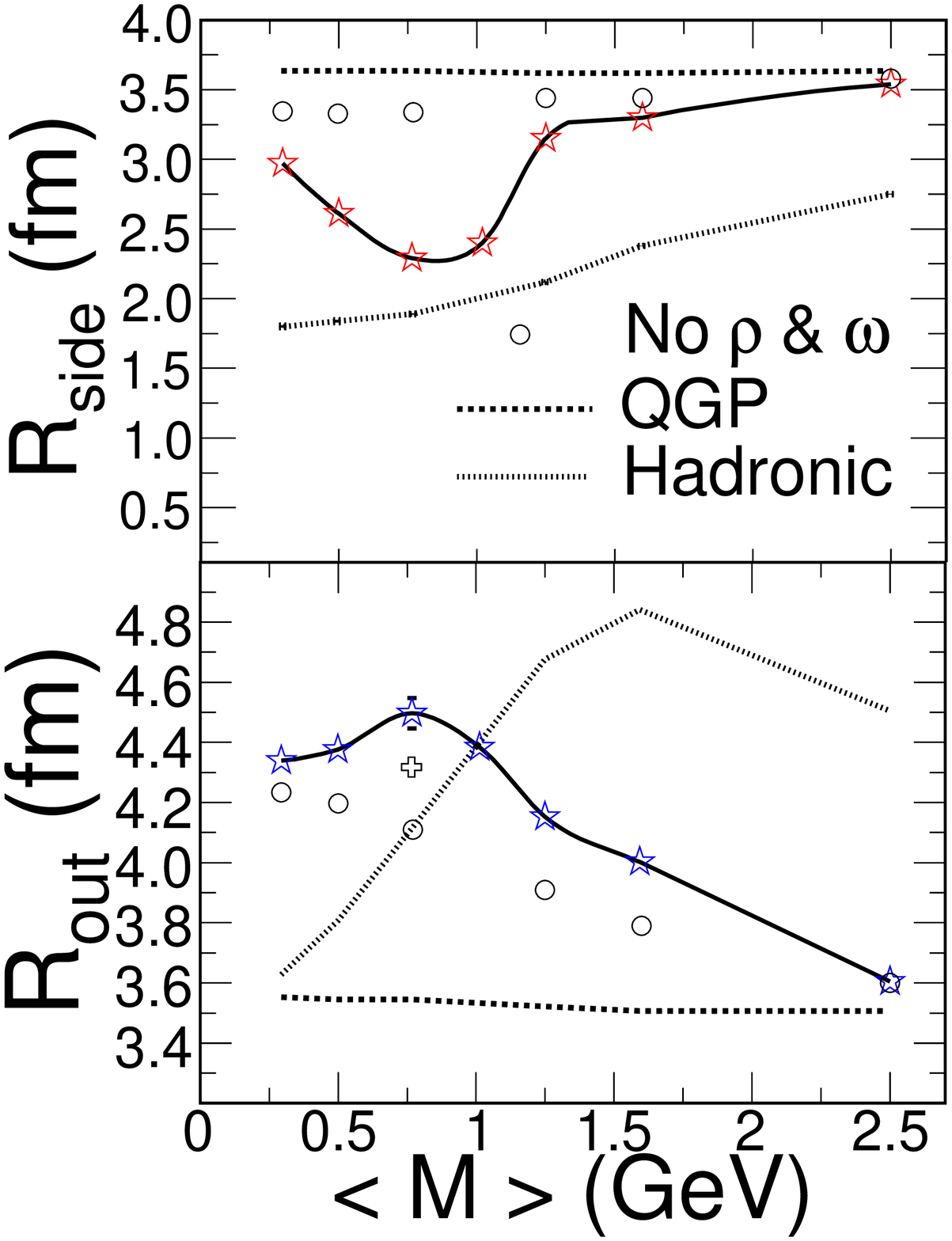}
\caption{(color online) $R_{side}$ and $R_{out}$ as a function of 
$\langle M \rangle$. The dashed, dotted and 
the solid line (with asterisk) indicate the HBT radii 
for the QGP, hadronic and
total dilepton contributions from all the phases respectively.
The open circles are obtained by 
switching off the contributions from $\rho$ and $\omega$. 
}
\label{fig3}
\end{center}
\eef
While the radius ($R_{\mathrm side}$) corresponding to $q_{side}$ 
is closely related to the transverse size of the system and 
considerably affected by the collectivity, 
the radius ($R_{\mathrm out}$) corresponding to $q_{out}$ measures both the 
transverse size and duration of particle emission~\cite{hb3,uaw}. 
The extracted $R_{\mathrm side}$ and $R_{\mathrm out}$ for different 
$\langle M\rangle$ are shown in Fig.~\ref{fig3}.
The $R_{\mathrm side}$ shows non-monotonic dependence on $M$.
The $R_{\mathrm side}$ reduces with $\langle M\rangle$, reaches its
minimum value at $\langle M\rangle\sim m_\rho$ and then increases 
again at high $\langle M\rangle$ approaching values close to  
the corresponding $R_{\mathrm side}$ for the QGP phase.
It can be shown
that $R_{side}\sim 1/(1+E_{\mathrm collective}/E_{\mathrm thermal})$.
In the absence of radial flow, $R_{\mathrm side}$ is independent of 
$M$. With the radial expansion of the system
a rarefaction wave moves
toward the center of the cylindrical geometry as a consequence the radial
size of the emission zone decreases with time. 
Therefore, the size of the emission zone is larger at early times   
and smaller at late time. 
The high $\langle M\rangle$ regions 
are dominated by the early partonic phase 
where the collective flow has not been developed fully 
{\it i.e.} the ratio of collective to thermal energy is small
hence show larger $R_{\mathrm side}$ for the source.
In contrast, the lepton pairs with $M\sim m_\rho$ 
are emitted from the late hadronic phase where the size of the emission zone 
and hence the $R_{\mathrm side}$ 
is smaller due to larger collective flow.
The ratio of collective to thermal 
energy for such cases is quite large, which is reflected as a dip
in the variation of $R_{\mathrm side}$ with $\langle M\rangle$ 
around the $\rho$-mass region (Fig.~\ref{fig3} upper panel). 
Thus the variation of $R_{\mathrm side}$
with $M$ can be used as an efficient tool to measure the
collectivity in various phases of the matter. 
The dip in $R_{\mathrm side}$ at $\langle M\rangle\sim
m_\rho$ is due to the contribution dominantly from the hadronic phase.
The dip, in fact vanishes if the contributions from $\rho$ and $\omega$ 
are switched off (circle in Fig.~\ref{fig3}). 
We observe that by keeping the $\rho$ and $\omega$ contributions
and setting radial velocity, $v_r=0$, the dip in $R_{\mathrm side}$
vanishes, confirming
the fact that the dip is caused by the radial flow of the hadronic matter.   
Therefore, the value of 
$R_{\mathrm side}$ at $\langle M\rangle\sim m_\rho$ may be used
to estimate the average $v_r$ in the hadronic phase.

The $R_{\mathrm out}$ probes both the transverse dimension and the 
duration of emission. Therefore, unlike $R_{\mathrm side}$, it does not
remain constant even in the absence of radial flow. The large $M$ regions are
populated by lepton pairs from early partonic phase where the
effect of flow is small and the duration of emission is also
small - resulting in smaller values of $R_{\mathrm out}$. 
For lepton pair from $M\sim m_\rho$ the flow is large
which could have resulted in a dip as in $R_{\mathrm side}$ in
this $M$ region. However, $R_{\mathrm out}$ probes the duration 
of emission too which is large for hadronic phase.
The larger duration compensates the reduction of $R_{\mathrm out}$ 
due to flow in the hadronic phase 
resulting in a bump in $R_{\mathrm out}$ in this region of $M$
(Fig.~\ref{fig3} lower panel). Both $R_{\mathrm side}$ and $R_{\mathrm out}$
approach QGP values for $\langle M\rangle\sim 2.5$ GeV implying dominant
contributions from the partonic phase in this region of $M$.

\bef
\begin{center}
\includegraphics[scale=0.3]{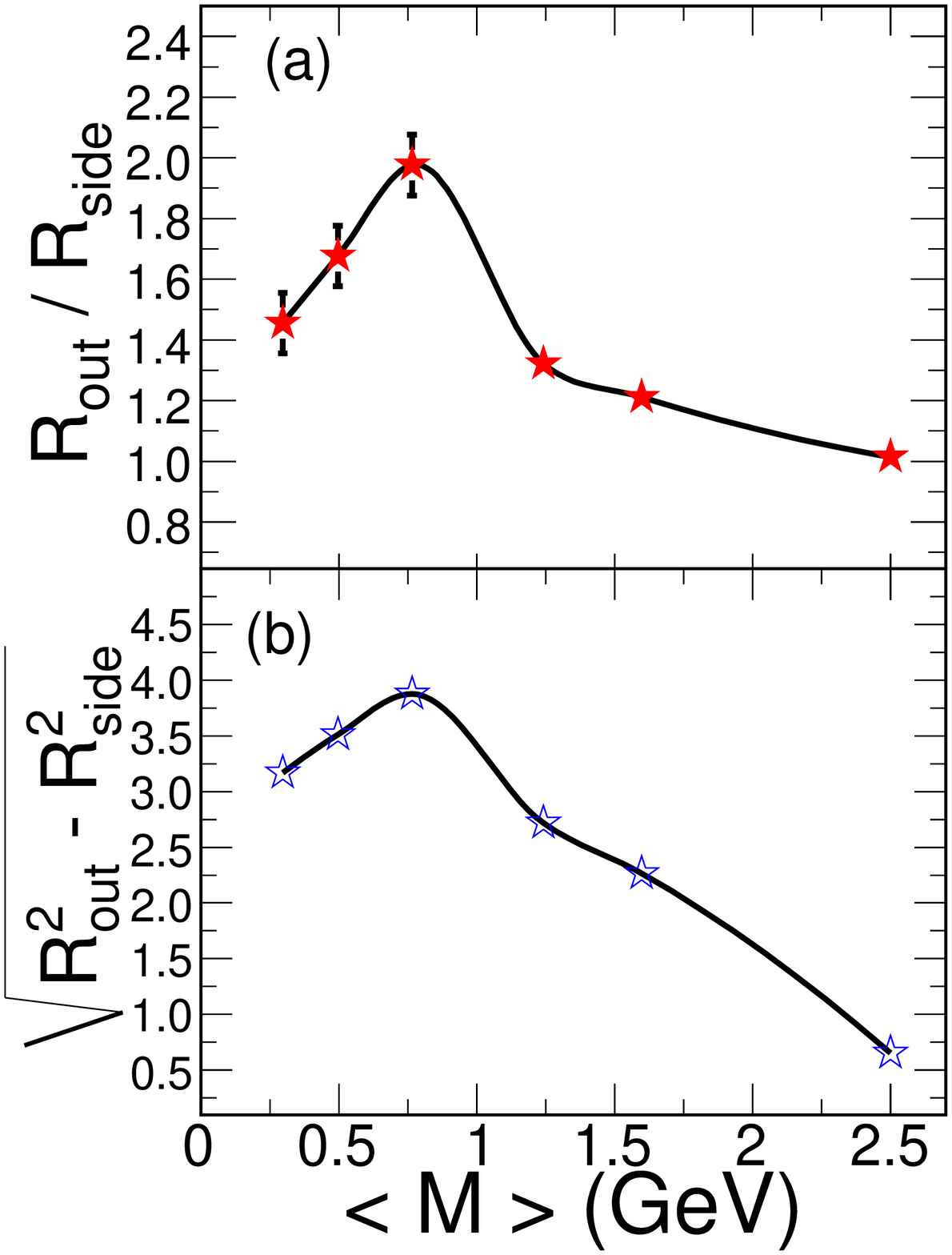}
\caption{(color online) The ratio $R_{out}$/$R_{side}$ and the difference $\sqrt{R_{out}^{2} - R_{side}^{2}}$ as a function of  $\langle M \rangle$.}
\label{fig4}
\end{center}
\eef

It is clear from the results displayed in Fig.~\ref{fig2} that
the high $M(>m_\phi$) and low $M(<0.5$ GeV) regions are dominated
by the radiation from early QGP phase when collective motion is not fully 
developed - as a result the size of the emission zones for these $M$
regions are large. However, lepton pairs for $M$ around $\rho$ mass dominantly
originate from the late hadronic phase when the collectivity is large and
consequently the size of the homogeneous emission domain is small.
This results in non-monotonic variation of $R_{\mathrm side}$ 
with $M$ as discussed above.  
Therefore, such a non-monotonic behaviour of the HBT radii
will signal the presence two different phases during the evolution - indicating
the creation of QGP which will inevitably reverts to hadrons 
because of the cooling due to expansion.  

The quantities $R_{\mathrm out}$ and $R_{\mathrm side}$ 
are proportional to the average size of the system~\cite{rischke}.
However, in the
ratio, $R_{\mathrm out}/R_{\mathrm side}$  some of the uncertainties 
associated with the space time evolution 
get canceled out. The quantity, $R_{\mathrm out}/R_{\mathrm side}$  
gives the duration of particle emission~\cite{hermm,chappm,hb11} 
for various domains of $M$.

Figure~\ref{fig4} shows the variation of the ratio, $R_{out}$/$R_{side}$ and the difference,
$\sqrt{R_{out}^{2} - R_{side}^{2}}$ as a function of  $\langle M \rangle$  
for Au+Au collisions at $\sqrt{s_{NN}}$ = 200 GeV. Both show a non-monotonic
dependence on $\langle M \rangle$.  The smaller values of both the quantities,
particularly at high mass region, 
reflect the contributions from the early partonic phase 
of the system. The peak around $\rho$-meson mass reflects dominance of
the contribution from the late hadronic phase as discussed before.

Now we discuss below two experimental challenges in such studies. 
We quote some numbers  
from the PHENIX measurements, keeping in mind that the
situation will further improved by increasing the luminosity
as well as collision energy.
The number of events can be computed from the luminosity (${\cal L}$),
the $pp$ in-elastic cross-section ($\sigma$) and the run time (${\cal T}$) of
the machine as, 
\begin{equation}
N_{\mathrm event}={{\cal L}}\times \sigma\times {\cal T}
\end{equation}
For RHIC, the luminosity, ${\cal L}$ is of the order of 
$50 \times 10^{27}$\,cm$^{-2}$\,sec$^{-1}$ and $\sigma$=40 mb. If RHIC runs for 
12 weeks then the number
of events, $N_{\mathrm event}=1.45\times 10^{10}$.
For the $M$,  $810\leq M$(MeV)$\leq 990$, the differential number 
($dN/2\pi p_Tdp_Tdy$) measured by PHENIX collaboration in Au+Au collisions
~\cite{phenixdil} at 
$\sqrt{s_{\mathrm NN}}=200$ GeV is give by:
\begin{equation}
\frac{dN}{2\pi p_Tdp_Tdy}\mid{y=0}=\frac{N_{part}}{2}\times 1.29\times 10^{-7}
\end{equation}
in the $p_T$ bin 1-1.8 GeV. 
Therefore, the (differential) number of pairs in the above range 
of $p_T$ and $M$ is $\sim 1.45\times 10^{10}\times \frac{N_{part}}{2}
\times 1.29\times 10^{-7}\sim\frac{N_{part}}{2}\times 1870$.

Similarly for $M$,   $500\leq$ M(MeV)$\leq 750$, the measured value of the
above quantity is:
\begin{equation}
\frac{dN}{2\pi p_Tdp_Tdy}\mid_{y=0}=\frac{N_{part}}{2}\times 2.235\times 10^{-7}
\end{equation}
for the $p_T$ bin 1.4-1.8 GeV. 
This indicates that the (differential) number of pairs in this kinematic
domain  is $\sim 1.45\times 10^{10}\times \frac{N_{part}}{2}
\times 2.235\times 10^{-7}\sim \frac{N_{part}}{2}\times 3240$.

For $0-10\%$ centrality the number of participants for Au+Au collisions
at $\sqrt{s_{\mathrm NN}}=200$ GeV is about 330. The number
of lepton pairs in the $p_T$ range 1.4-1.5 GeV is $\sim 5.3\times 10^{5}$
for the $M$ window 0.5-0.75 GeV.  
For 12 weeks of run time the number of events
estimated with the current RHIC luminosity is $\sim1.45\times 10^{10}$.
Then the number of pairs produced per event  is 
$\sim 3\times10^{-5}$ in the kinematic range 
mentioned above. The probability to have 2 pairs of dileptons is $\sim 
10^{-9}$. Therefore, roughly $10^9$ events are required 
to make the HBT interferometry with lepton pairs possible.

It is  expected that further increase in luminosity
at RHIC by a factor 2 beyond the year 2012 to about 
$10^{29}$ cm$^{-2}$ s$^{-1}$ may be a motivating factor for  
such measurements.  The increase in lepton pair production
at higher collision energy at Large Hadron Collider (LHC) may also 
provide a reason to pursue such measurements.

The possibility of dilution of signal due to addition
of random pairs, which one may encounter in the analysis of 
experimental data is discussed below. 
We have added some `` mixture'' to the dilepton source with exponential
energy distribution, {\it i.e.} we have replaced $\omega$ by 
$\omega+\delta\omega$ where $\delta\omega$ has exponential energy (of the
pair) dependence and  weight factor is as large as that of $\omega$ itself.
Then we find that
the resulting change in the  HBT radii is negligibly small. This
can be understood from the expression for $C_2$ (Eq.~\ref{eq1}): 
\begin{eqnarray*}
C_2&&=1+\frac{\int d^4x_1\omega(x_1,K)\cos(\alpha_1)\int d^4x_2 
\omega(x_2,K)\cos(\alpha_2)} 
{\int d^{4}x\omega (x,\vec{k_1}) \int d^{4}x\omega (x,\vec{k_2})}\nn\\
&&+\frac{\int d^4x_1\omega(x_1,K)\sin(\alpha_1)
\int d^4x_2\omega(x_2,K)\sin(\alpha_2)]}
{\int d^{4}x\omega (x,\vec{k_1}) \int d^{4}x\omega (x,\vec{k_2})}
\end{eqnarray*}
where
\begin{widetext}
\begin{eqnarray*}
\alpha_1&=&\tau_1M_{1T}\cosh(y_1-\eta_1)-r_1k_{1T}
\cos(\theta_1-\psi_1)-\tau_1M_{2T}\cosh(y_2-\eta_1)+r_1k_{2T}
\cos(\theta_1-\psi_2)\nonumber\\
\alpha_2&=&\tau_2M_{2T}\cosh(y_2-\eta_2)-r_2k_{2T}
\cos(\theta_2-\psi_2)-\tau_2M_{1T}
\cosh(y_1-\eta_2)+r_2k_{1T}\cos(\theta_2-\psi_1)
\end{eqnarray*}
\end{widetext}
where $x_i=(\tau_icosh\eta_i, r_{iT}cos\theta_i, r_{iT}cos\theta_i,
\tau_isinh\eta_i)$. 

It is clear that the expression for $C_2$ contains quadratic 
power of the  source function
both in the numerator and denominator. Therefore, changes
in the source function will  lead to some sort of partial 
cancellation (complete
cancellation is not possible because the source function appears in the
numerator and the denominator inside the integral with different  
dependent variables, $K$ or $k_i$'s). 

Now at this point some comments on the change of the spectral function 
of light vector mesons in the medium are in order here.  
The modification of the spectral function of light vector mesons at non-zero
temperatures and densities is a field of high contemporary interest
both from the experimental  and theoretical 
(see ~\cite{ja2,rapp,brown} for reviews) point of views. The enhanced 
production of lepton pairs at the low $M$ domain in In+In collisions
at 158 A GeV beam energy  measured by NA60 collaboration~\cite{NA60}
indicates substantial broadening of the $\rho$ spectral function.  
In the present work 
we artificially modify the spectral function of the $\rho$ by increasing
its width by a factor of two  and keeping the pole mass at its 
vacuum value and then evaluate $C_2$ with the modified spectral function
to explicitly check the change in the magnitudes of the HBT radii.  
We observe that an increase in the width of $\rho$ by a factor of two
leads to the reduction of $R_{\mathrm side}$ by about $10\%$ 
for invariant mass windows below $\rho$-peak. This reduction is 
due to the contributions from the decays of broader $\rho$
which undergoes larger flow compared to the QGP phase.  
The nature of the variation of $R_{\mathrm side}$ with $\langle M\rangle$
remains unaltered. However, a huge broadening of $\rho$ for which the
contribution from the hadronic phase becomes overwhelmingly large
compare to the QGP phase may alter the nature of the variation 
of $R_{\mathrm side}$ in the low $M$ domain.  

\section{Summary}
In summary, the dilepton pair correlation functions has been evaluated for
Au+Au collisions at RHIC energy. The additional kinematic 
variable, $M$ for  dilepton pairs make it a more useful tool for characterizing
the different phases of the matter formed in heavy ion collisions 
compared to the HBT interferometry with direct photons.
The HBT radii extracted from the dilepton correlation 
functions show non-monotonic dependence on the invariant mass,
reflecting the evolution of collective flow in the system which
can be considered as a signal for the QGP formation in  
heavy ion collisions.  The $M$ dependence of the 
$R_{out}/R_{side}$ and $\sqrt{R_{out}^{2} - R_{side}^{2}}$ 
which can be experimentally measured could be used to characterize
the source properties at various stages of the evolution.

{\bf Acknowledgment:} We are grateful to Tetsufumi Hirano for providing
us the hadronic chemical potentials. 
We thank Nu Xu for very useful discussions. 
J A and P M supported by DAE-BRNS project Sanction No.
2005/21/5-BRNS/2455.

\end{document}